\documentclass[conference]{IEEEtran}

\usepackage{cite}
\usepackage{graphicx}
\usepackage[tight,footnotesize]{subfigure}
\usepackage{pifont}
\usepackage{amssymb}
\usepackage{dblfloatfix}
\usepackage{booktabs}
\usepackage{amsmath}
\usepackage{upgreek}

\begin{document}
\bstctlcite{IEEEexample:BSTcontrol}
\title{Cryogenic Characterization of 28\,nm Bulk CMOS Technology for Quantum\,Computing\vspace{-0.7cm}}
\author{\IEEEauthorblockN{Arnout Beckers\IEEEauthorrefmark{2},
Farzan Jazaeri\IEEEauthorrefmark{2},
Andrea Ruffino\IEEEauthorrefmark{3}, 
Claudio Bruschini\IEEEauthorrefmark{2}\IEEEauthorrefmark{3}, Andrea Baschirotto\IEEEauthorrefmark{4}, and
Christian Enz\IEEEauthorrefmark{2}}
\IEEEauthorblockA{\IEEEauthorrefmark{2}Integrated Circuits Laboratory (ICLAB), Ecole Polytechnique F\'ed\'erale de Lausanne (EPFL), Switzerland,\\
\IEEEauthorrefmark{3}Advanced Quantum Architecture Lab. (AQUA), Ecole Polytechnique F\'ed\'erale de Lausanne (EPFL), Switzerland,\\
\IEEEauthorrefmark{4}INFN \& University of Milano-Bicocca, Milano, Italy,\\}
arnout.beckers@epfl.ch
\vspace{0.4cm}}

\maketitle

\begin{abstract}
This paper presents the first experimental investigation and physical discussion of the cryogenic behavior of a commercial 28\,nm bulk CMOS technology. Here we extract the fundamental physical parameters of this technology at 300, 77 and 4.2\,K based on DC measurement results. The extracted values are then used to demonstrate the impact of cryogenic temperatures on the essential analog design parameters. We find that the simplified charge-based EKV model can accurately predict the cryogenic behavior. This represents a main step towards the design of analog/RF circuits integrated in an advanced bulk CMOS process and operating at cryogenic temperature for quantum computing control systems. 
\end{abstract}
%\begin{keywords}
%\normalfont{\textbf{28\,nm bulk CMOS, EKV,  slope factor, quantum computing, cryogenic, 4.2\,K}}
%\end{keywords}
\IEEEpeerreviewmaketitle
\let\thefootnote\relax\footnote{\hspace*{-1em}This project has received funding from the European Union's Horizon 2020 Research \& Innovation Programme under grant agreement No. 688539 {MOS-Quito}. The 28\,nm bulk technology was provided by ScalTech28, INFN Milano-Bicocca.}

\section{Introduction}
Quantum computing promises a rapid enhancement of the available computational power for selected algorithms while transistor scaling is slowing down. Not long ago, it has been proposed that quantum bits (``qubits") can be implemented in the electron spins in silicon~\cite{pla} and in conformity with industry-standard CMOS technology~\cite{leti}. This triggered growing interest in the co-integration of the qubits and their control system, requiring the development of dedicated analog/RF CMOS electronics for the initialization, manipulation and read-out of the qubits~\cite{charb}. However, the theoretical exponential increase of the computational power is only accessible if the qubits'\,entangled superpositions can be preserved during computation. This is achieved in practice by operating the qubits at deep cryogenic temperature (mK-range), which reduces the thermal noise. In a co-integrated system also the control system operates at cryogenic temperature, which can further reduce the thermal noise by removing the need for direct interconnections from a room temperature (RT) control system to the qubits. In\cite{leti} the silicon qubits are implemented on silicon oxide and operate at mK-temperature. Nonetheless, the exact operating conditions of the control system are still unclear to date. For instance, a 3D-integrated system with a temperature gradient from top to bottom~\cite[Fig.~13]{charb}, where the control system is at a higher cryogenic temperature (e.g.\,77\,K, 4.2\,K) than the qubits, may be a better solution in terms of analog performance, heat removal or various non-thermal noise sources. In a later stage, co-integration also has the advantage of scalability to large qubit arrays. Compact transistor models extended to cryogenic temperatures will then be a must to increase the chance of first silicon right quantum computing systems. As we will bring forward as the key finding of this work, the simplified charge-based EKV model presents an interesting first step towards the development of such a cryogenic compact model for advanced technology nodes. 

\begin{table}
	\centering
	\caption{Measured devices (28\,nm Bulk CMOS Process)}
	\label{table}
	\resizebox{0.24\textwidth}{!}{%
\begin{tabular}{rcc}
	\toprule
	Symbol & Type & W/L\\ 
	\midrule
{\tiny\ding{108}} & $n$MOS & 3\,$\upmu$m / 1\,$\upmu$m\\
	{\tiny\ding{115}} & $p$MOS & 3\,$\upmu$m / 1\,$\upmu$m\\
	{\tiny\ding{110}} & $n$MOS & 1\,$\upmu$m / 90\,nm\\
	{\tiny\ding{116}} & $n$MOS & 3\,$\upmu$m / 28\,nm\\
	{\tiny\ding{117}} & $n$MOS & 300\,nm / 28\,nm\\ 
	\bottomrule
\end{tabular}}

\end{table}

\begin{figure*}[t]
	\centering
	\includegraphics[scale=0.9]{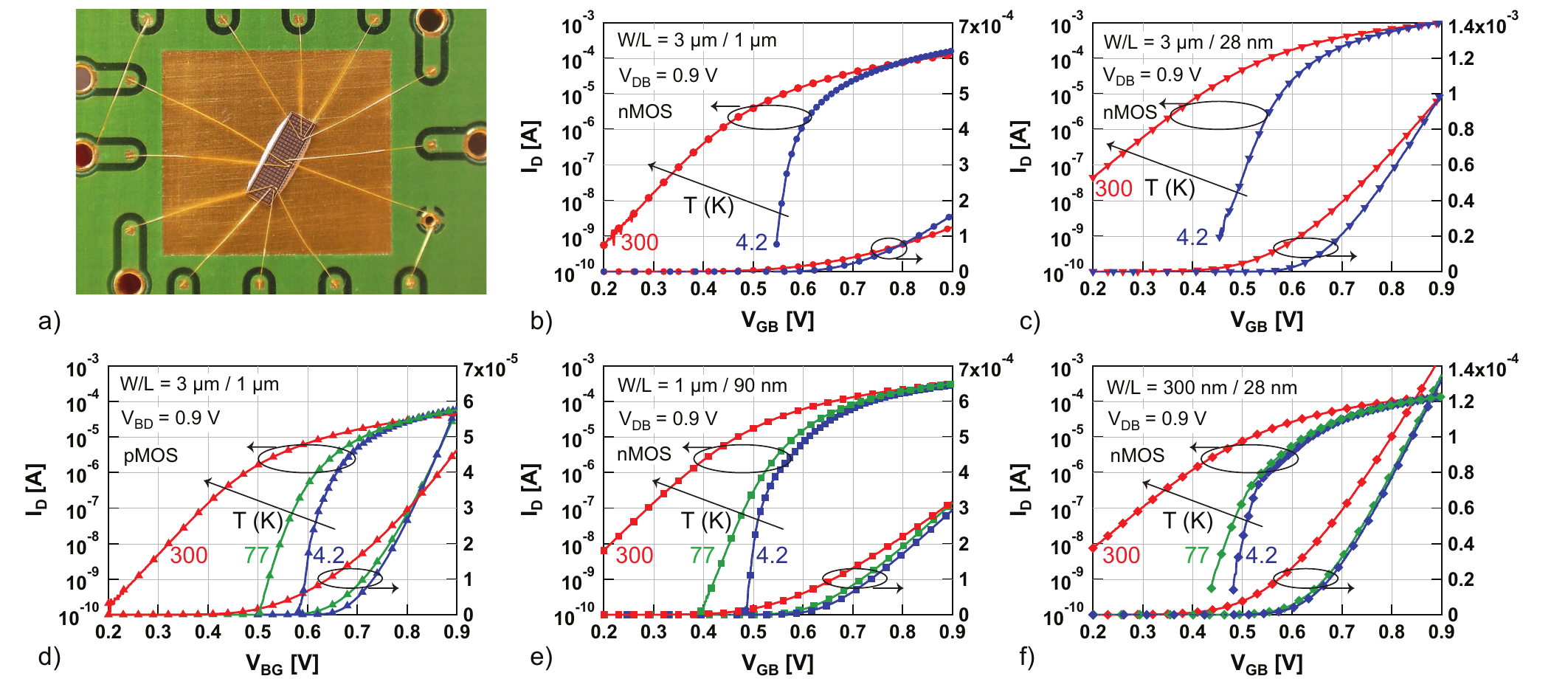}
	\hfil
	\vspace{-0.65cm}
	\caption{Cryogenic measurements on a 28\,nm bulk CMOS technology, a) Au-wire bonded sample chip glued to a standard PCB, before covering with a globtop, b)-f) Measured transfer characteristics at 300, 77 and 4.2\,K in saturation  ($V_{DB}$ = 0.9 V). In all measurements the gate voltage is referred to the bulk ($V_{GB}$) and swept from 0.2\,V to 0.9\,V with a step size of 1\,mV. Marker symbols refer to the device type (as shown in table \ref{table}). Colors indicate the temperature, red: room temperature (300\,K), green: liquid nitrogen temperature (77\,K) and blue: liquid helium temperature (4.2\,K).}\label{fig:1}
\end{figure*}

\section{Cryogenic CMOS Electronics}
The control system involves the following analog/RF building blocks, which are to be designed in cryogenic CMOS electronics (``cryo-CMOS''): multiplexers to control multiple qubits at once, low-noise amplifiers and oscillators. For the design of these building blocks, the advanced technology nodes (below 100\,nm) are the most important because of their potential for co-integration, low bias operation and very high transit frequency $F_t$, reaching several hundreds of GHz even at RT. This high $F_t$ can be traded with power consumption by shifting the bias point to weak inversion, where $F_t$ reaches tens of GHz, which is still high enough for qubit manipulation and read-out~\cite{leti}. On top of that, at cryogenic temperatures we expect to have an additional increase in $F_t$ in weak inversion, allowing for even more current savings. 

Previous research shows cryogenic measurement results for the following advanced technology nodes: a 40\,nm bulk CMOS process at liquid helium temperature (4.2\,K)~\cite{charb} and a 28\,nm FDSOI process at liquid nitrogen temperature (77\,K)~\cite{shin}. In this work, based on cryogenic measurement results of a 28\,nm bulk CMOS technology at 77\,K and 4.2\,K, we quantify the impact of cryogenic temperature on the essential analog design parameters, namely i) the transconductance $G_m$ and transconductance efficiency $G_m / I_D$ used for low power analog design; ii) the intrinsic gain $G_m / G_{ds}$, important for amplifying the weak signals coming from the qubits, and iii) the transit frequency $F_t$. This will allow to design low power analog/RF control building blocks that work properly at cryogenic temperature and have a minimal effect on the qubits. 

\section{Cryogenic Measurements}
\subsection{Measurement Set-up} The measurements were conducted on the devices presented in Table I, fabricated in a 28\,nm bulk CMOS process~\cite{zhang, pezzotta}. The sample chips were first wire bonded to standard PCBs using Au-wire bonds (Fig.~1a) and then covered with a glob-top. The PCBs were immersed into liquid nitrogen (77\,K) and liquid helium (4.2\,K) by means of a dipstick. The results were acquired with a Keysight B1500A semiconductor device parameter analyzer. Using this set-up, we measured transfer characteristics in the linear ($V_{DB}$ = 10\,mV) and the saturation regions ($V_{DB}$ = 0.9\,V), as well as output characteristics for different gate voltages. 
\begin{figure*}[t]
	\includegraphics[scale=0.84]{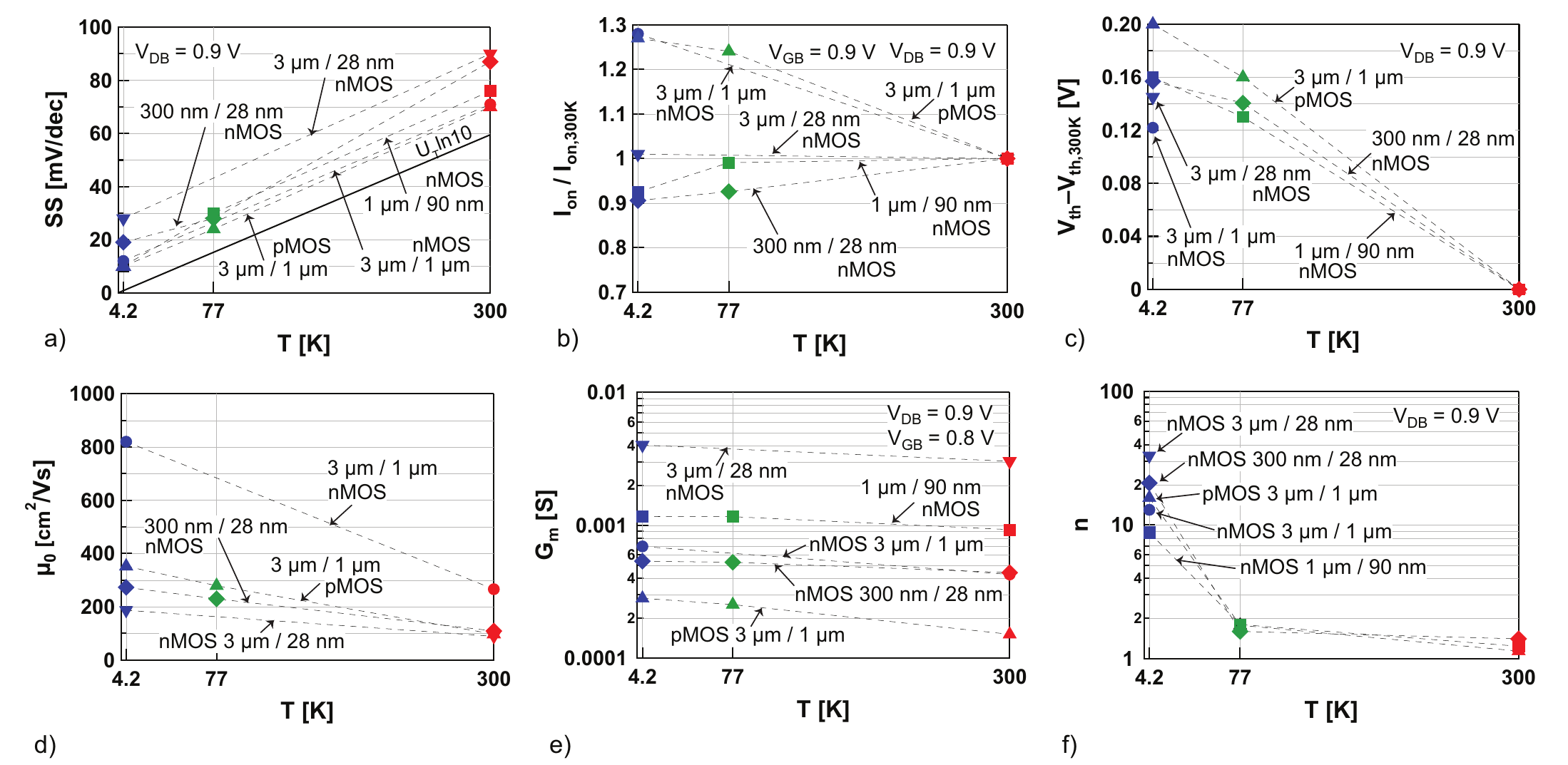}
	\hfil
    \vspace{-0.68cm}
	\caption{Extraction of the fundamental physical and technology parameters at 300, 77 and 4.2\,K, a) Subthreshold swing versus temperature. The results at 4.2\,K show a strong deviation from the theoretical trend ($U_T\ln10$), which is expressed by an increase in the slope factor $n$, b) On-state current normalized to RT, c) Shift in threshold voltage at 77\,K and 4.2\,K with respect to RT, extracted from the transconductance in saturation ($V_{DB}$\,=\,0.9 V) at $V_{GB}$\,=\,0.9\,V, d) Low field mobility versus temperature, extracted using the $Y$-function method~\cite{ghibaudo}, e) Transconductance in saturation versus temperature, f) Slope factor versus temperature.}\label{fig:2}
\end{figure*}
\subsection{Measurement Results and Discussion}
Figs.~1b-f show the measured transfer characteristics in saturation for the devices in Table~\ref{table}. Clearly, the subthreshold swing $SS$, defined as $nU_T\ln10$ [mV/dec] with $U_T \triangleq kT/q$ the thermal voltage, decreases drastically at 77\,K and 4.2\,K for all devices. The slope factor $n$ expresses the deviation of the $SS$ from the asymptote corresponding to an ideal device with $n$\,=\,1. As illustrated in Fig.~2a, the $SS$ decreases to 11\,mV/dec at 4.2\,K ($-$85\,\%) for long channel $n$MOS $(L$\,=\,1\,$\upmu$m$)$, although this decrease is less than the expected value from the ideal case corresponding to $\approx$\,0.8 mV/dec, pointing out the much higher slope factor $n$ than unity at 4.2\,K. This is attributed to the incomplete dopant ionization at cryogenic temperatures. In short channel $n$MOS ($L$\,=\,28\,nm), the $SS$ reaches 28 mV/dec ($-$68\,\%). The on-state current decreases at 77\,K and 4.2\,K for the short devices, while it increases for the long devices (Fig.~2b). This can be explained from the expression of the (electron) current density $\emph{\textbf{J}}_n$\,=\,$qn\mu_n\emph{\textbf{E}} + qD_n\nabla n$ with $D_n=\mu _n U_T$. At low bias, $J_n$ is proportional to the temperature and field-assisted ionization is weak, whereas at high bias the current density increases with decreasing temperature through the mobility and field-assisted ionization is strong in the channel~\cite{cor}. 
\begin{figure*}[t]
	\centering
	\includegraphics[scale=0.87]{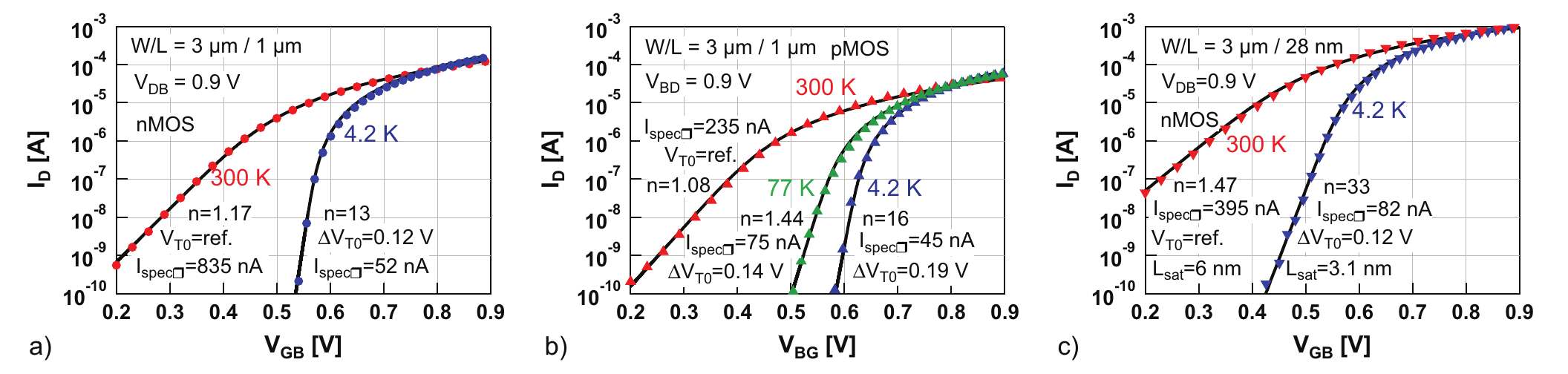}
	\hfil
	\vspace{-0.35cm}
	\caption{Transfer characteristics: measured (markers) and modeled (solid lines) with the simplified EKV long/short channel model  for a) $n$MOS $W/L$\,=\,3 $\upmu$m\,/\,1\,$\upmu$m  at RT and 4.2\,K, b) $p$MOS $W/L$\,=\,3 $\upmu$m\,/\,1\,$\upmu$m at RT, 77\,K and 4.2\,K and c) $n$MOS $W/L$\,=\,3\,$\upmu$m\,/\,28\,nm at RT and 4.2\,K. For each curve, the extracted model parameters are shown.}\label{fig:4}
\end{figure*} 
Additionally, the threshold voltage $V_{th}$ shifts to higher voltages at 77\,K and 4.2\,K due to incomplete ionization. Indeed, a higher voltage is needed to attract sufficient charges to the surface to create the inversion layer. The shift in threshold voltage $\Delta V_{th}$ with respect to RT in Fig.~2c was extracted from the transconductance in saturation and shows a significant increase in the order of 0.1\,V. It should be noted that the maximum threshold voltage variation is observed in long channel $p$MOS devices $(\Delta V_{th}$\,=\,0.2\,V). Moreover, relying on the $Y$-function method~\cite{ghibaudo} the low-field mobility $(\upmu_0)$ was extracted for different gate lengths. Fig.~2d shows a significant improvement ($\times$\,3 for $n$MOS, $L$\,=\,1\,$\upmu$m) at 4.2\,K due to the phonon scattering reduction, although the Coulomb scattering is becoming dominant at low temperature. The transconductance in saturation $G_{m,sat}$ improves at 4.2\,K ($\times$\,1.3 for $n$MOS, $W/L$\,=3\,$\upmu$m/\,28\,nm). Fig.~2f shows that at 77\,K the slope factor $n$ remains close to the value at RT, e.g. 1.6 compared to 1.47 for $n$MOS $W/L$\,=300\,nm/\,28\,nm. This slight increase at 77 K was also observed in~\cite{bink}. Here we report for the first time on the strong increase of $n$ at 4.2\,K (Fig.~2f), which was evident from the extracted values of the $SS$ in Fig\,2a. The extraction procedure of the slope factor is demonstrated in Fig.~4a at 300\,K and 4.2\,K. For $n$ = 33 ($n$MOS, $W/L$\,=3\,$\upmu$m/\,28\,nm), the modified theoretical trend $nU_T\ln10$ indicates a value of 27.5 mV/dec, in accordance with the measured value in Fig.~2a. 
\begin{figure*}[!t]
	\includegraphics[scale=0.84]{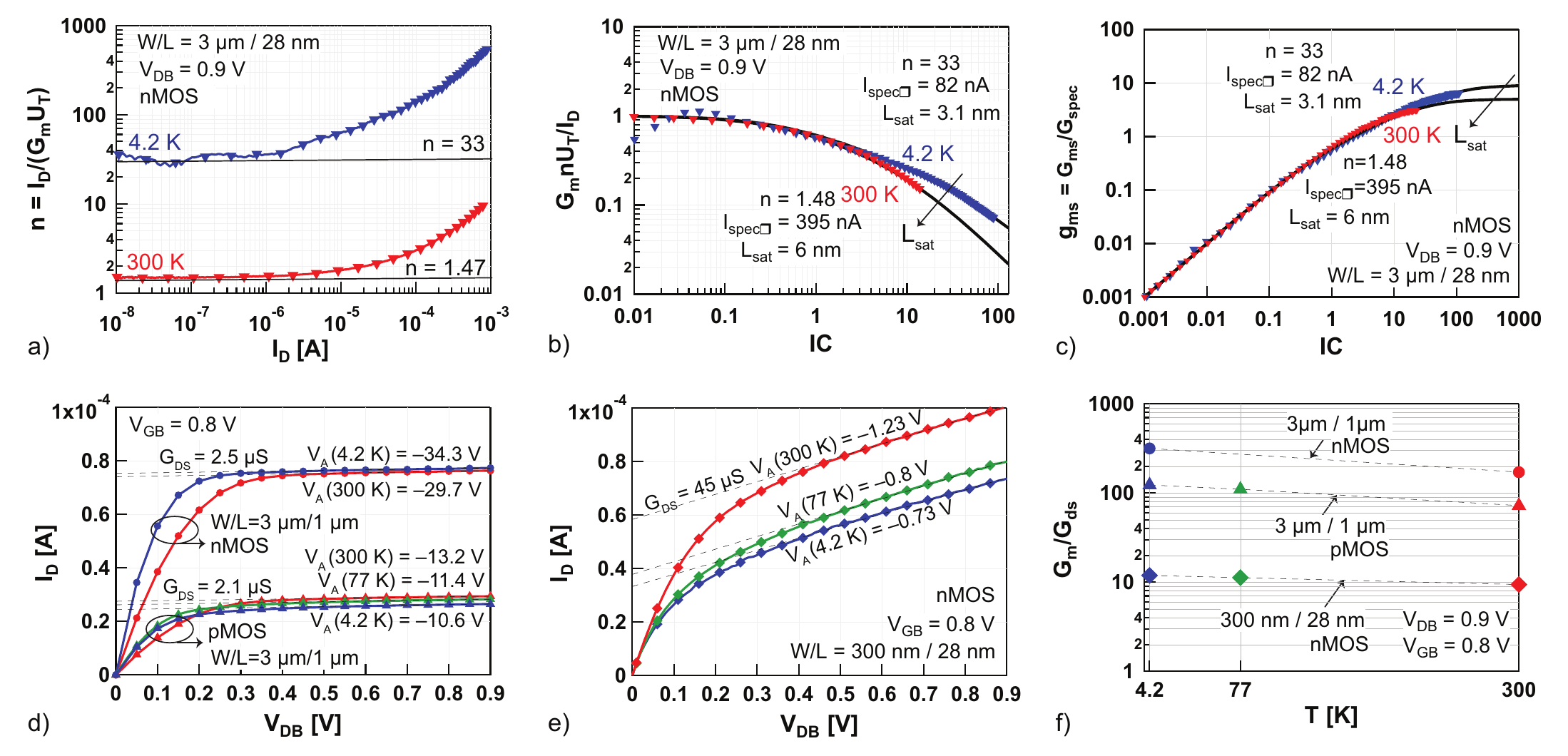}
	\hfil
	\vspace{-0.68cm}
	\caption{Impact of cryogenic temperatures on analog design parameters, a) Slope factor versus the drain current at \,300\,K and 4.2\,K for $n$MOS $W/L$\,=\,3\,$\upmu$m\,/\,28\,nm, b) Normalized transconductance efficiency versus the inversion coefficient for $n$MOS $W/L$\,=\,3\,$\upmu$m\,/\,28\,nm, showing a decreased velocity saturation effect at 4.2\,K. Solid lines: model, c) Normalized source transconductance versus the inversion coefficient. Solid lines: model, d) Measured output characteristics for long channel $n$MOS and $p$MOS $W/L$\,=\,3\,$\upmu$m\,/\,1\,$\upmu$m with extracted values for the output conductance  and the Early voltage $V_A$, e) Measured output characteristics for short channel $n$MOS $W/L$\,=\,300\,nm\,/\,28\,nm with extracted values for the output conductance and $V_A$, f) Intrinsic gain at 300, 77 and 4.2\,K.}\label{fig:3}
\end{figure*} 
\section{Cryogenic Analog Design Parameters}
To extract the analog design parameters, we use the simplified and normalized EKV model, described in detail in ~\cite{mangla}. This model captures all the changes of the temperature reduction to cryogenic temperatures in only four parameters. The long channel model relies on three model parameters, i.e. the slope factor $n$, the threshold voltage $V_{T0}$  and the specific current ($I_{spec}=I_{spec\scriptsize{\square}}W/L=2(W/L)n\mu C_{ox}U_T^2$). The current is normalized to $I_{spec}$ to obtain the inversion coefficient $IC \triangleq I_{D,sat}/I_{spec}$. Initial guesses for the model parameters can be estimated from the extracted values. The short channel model also adds the saturation length $L_{sat}$ as a fourth model parameter, indicating the part of the channel in full velocity saturation. As shown in Fig.~3, both the RT and cryogenic measurements can be accurately predicted by the proposed model for long (Figs.~3a-b) and short (Fig.~3c) devices. It is worth mentioning that the model parameters obtained here confirm the extracted results from Figs.~2c~and~2f. Furthermore, $I_{spec}$ decreases by one order of magnitude from RT to 4.2\,K due to its quadratic dependency on $U_T$ in the model, which is explained physically by incomplete ionization. In Figs.~4b-f the impact of cryogenic temperatures on the essential analog design parameters is analyzed. The transconductance efficiency $G_mnU_T/I_D$ in Fig.~4b and the normalized transconductance at the source $g_{ms}=nG_{m}/G_{spec}$ with $G_{spec}=I_{spec}/U_T$ in Fig.~4c, both show a lower impact of velocity saturation at 4.2\,K in strong inversion ($IC > 10$). The saturation length $L_{sat}$ is reduced from 6\,nm at RT to 3\,nm at 4.2\,K $(L$\,=\,28\,nm$)$. As can be seen in the output characteristics plotted in Fig.~4d for long channel and in Fig.~4e for short channel, the output conductance $G_{ds}$ remains practically constant with respect to temperature. 
%because the transistor carries less charges at a higher mobility. 
This results in an increased intrinsic gain $G_m/G_{ds}$ at 77\,K ($\times$\,1.2) and 4.2\,K ($\times$\,1.3, for $n$MOS, $W/L$\,=300\,nm/\,28\,nm), which is promising for cryogenic amplifier design (Fig.~4f). Assuming the capacitances do not change significantly going down in temperature~\cite{akturkcapacitance}, the transit frequency $F_t$ follows the increase in the transconductance shown in Fig.~2e. This increase can be traded-off for a lower power consumption, which is also beneficial in terms of heat dissipation from the control system to the qubits. The current saved at cryogenic temperature to reach the same $F_t$ as at RT, can be evaluated in weak inversion as 
\begin{equation*}
\frac{I_D}{I_{D0}}=\frac{n}{{n_0}}\cdot\frac{T_{cryo}[K]}{300}, \label{2} 
\end{equation*}  
where $n_0$ and $I_{D0}$ are the RT-values. Ideally, if $n$ would stay the same at RT as at 4.2\,K, the current reduction factor would be 71. Unfortunately, due to the strong increase of $n$ at 4.2\,K mentioned above, the current reduction is only 5.2 for $W/L$\,=\,300\,nm\,/\,28\,nm ($n$ = 20), and even 3.2 for $W/L$\,=\,3\,$\upmu$m\,/\,28\,nm ($n$ = 33) due to lower electrostatic control of wider channels. Since the change of $n$ at 77\,K is only minor (typically 1.6), the current reduction at 77\,K is a factor of 3.6. In other terms, the $n$-factor mitigates the expected current savings moving from 77\,K to 4.2\,K for reaching the same $F_t$.
\section{Conclusion}
This work presents the influence of cryogenic temperature on a 28\,nm bulk CMOS technology for quantum computing control systems. Starting from a detailed analysis of the physical device properties at cryogenic temperatures, encouraging trends in the essential analog design parameters are obtained, although the strong increase in the slope factor at 4.2\,K mitigates the expected current savings. The proposed analysis demonstrates, by means of the simplified EKV model, that the cryogenic behavior in advanced CMOS can be accurately predicted. This represents an interesting solution for further implementation in cryogenic compact models for silicon-based quantum computing systems.
\newpage
\section*{Acknowledgement}
The authors would like to thank G.~Boero and A.~Matheoud for sharing their expertise in cryogenic measurements, G.~Corradini for the wire-bonding and P.~Van~der~Wal for providing the liquid nitrogen (all EPFL). 

\bibliographystyle{ieeetran}
\bibliography{essderc}
\end{document}